\documentstyle[epsf,12pt]{article}
\newcommand{\yourabstract}[1]{
\mbox{}\\
\mbox{}\\
{\bf\noindent Abstract}\\
\begin{center}
\mbox{}\parbox[t]{5.in}{#1}
\end{center} }

\newcommand{\half}{\textstyle{\frac{1}{2}}}
\newcommand{\quarter}{\textstyle{\frac{1}{4}}}

\newcommand{\beq}{\begin{equation}}
\newcommand{\beqar}{\begin{eqnarray}}
\newcommand{\eeq}[1]{\label{#1} \end{equation}}
\newcommand{\eeqar}[1]{\label{#1} \end{eqnarray}}
\def\prd#1{{Phys. Rev. }{D#1} }
\def\prl#1{{Phys. Rev. Lett. }{#1} }
\def\pl#1{{Phys. Lett. }{#1} }
\def\np#1{{Nucl. Phys. }{#1} }

\def\rmp#1{{Rev. Mod. Phys. }{#1} }

\begin{document}
\begin{titlepage}
\begin{flushright}
CERN-TH.7372/94 \\
hep-ph/9407377 \\
July 1994
\end{flushright}
\vskip 0.7cm
\begin{center}
{\bf \large Vacuum-Induced Quantum Decoherence}
\vskip 0.2cm
{\bf \large and the Entropy Puzzle}
           \footnote{
           Work supported by the Heisenberg
           Programme (Deutsche Forschungsgemeinschaft).
           }
\vskip 0.6cm
Hans-Thomas Elze
\vskip 0.2cm
CERN-Theory,
CH-1211 Geneva 23,
Switzerland \footnote{E-mail address: ELZE@CERNVM.CERN.CH}
\end{center}
\yourabstract{Or: ``How to generate an ensemble
in a single event?'' Following recent work
on entropy in strong interactions, I
explain the concept of
environment-induced quantum decoherence in
elementary quantum mechanics.
The classically chaotic inverted oscillator
becomes partially decoherent already in the
environment of a single other oscillator performing
only vacuum fluctuations.
One finds exponential entropy growth in the
subsystem with a Lyapunov exponent, which
approaches the classical one for weak coupling.
}
\vskip 0.2cm
\begin{center}
{\it Presented at the Workshops on ``Quantum Infrared Physics'', Paris,
 6$-$10.6.94, and ``QCD 94'', Montpellier, 7$-$13.7.94.
 Invited talk at the NATO Adv. Res. Workshop on ``Hot Hadronic Matter'',
 Divonne, 27.6.$-$1.7.94. To appear in the Proceedings.}
\end{center}
\end{titlepage}

Recently the long-standing ``entropy puzzle'' of
high-multiplicity events in strong interactions at ultra-relativistic
energies has been analysed from a new point of view \cite{I}. This is
related to the concepts of an {\it open quantum system} and
{\it environment-induced quantum decoherence}.
The problem dates
back to Fermi and Landau
and is intimately connected to understanding the
rapid thermalization of high energy density ($\gg 1\;\mbox{GeV/fm}^3$)
matter \cite{Fermi}. Why do thermal models work so well in reproducing
global features of hadronic multiparticle final states? Why do they
work at all?

Or, Why does high-energy
scattering of pure initial states lend itself to a
statistical description characterized by a large apparent {\it entropy}
from a mixed-state density matrix describing
intermediate stages in a space-time picture of parton evolution?
Effectively, {\it unitary time evolution} of the observable part
of the system breaks down in the
transition from a quantum mechanically pure initial state to a
highly impure (more or less thermal) high-multiplicity final state.
Note that the unitary time evolution operator, $\exp (-i\hat{\mbox{H}}
t)$, always transforms a pure state into a pure state, according to the
Schr\"odinger equation, which {\it cannot} produce entropy under any
circumstances (cf. below).
This was discussed in detail in Refs. \cite{I}, where more
references concerning formal aspects of this work can be found.
Based on analogies with studies of the quantum measurement process
(``collapse of the wave function'') \cite{Zu0} and
motivated by related problems in quantum cosmology and by
non-unitary non-equilibrium evolution
resulting in string theory \cite{GH}, I argued that
{\bf environment-induced quantum decoherence solves the
entropy puzzle} of strong interactions.

A complex pure-state quantum system can show a
quasi-classical behaviour, i.e. an impure density (sub)matrix together
with decoherence of the associated pointer states
in an observable subsystem \cite{I,Zu0,GH}. I will demonstrate in
the following that the decoherence process is uniquely correlated
with entropy production. Considering strong interactions, in
particular, there is a natural {\it Momentum Space Mode Separation}
due to confinement, which is defined in
the frame of initial conditions for the time evolution and for
the physical (gauge) field degrees of freedom. Thus, almost constant
QCD field configurations form an {\it unobservable environment},
which interacts with the {\it observable subsystem} composed
of partons. The environment modes are unobservable, since
they can neither hadronize nor initiate hard scattering among themselves,
 whereas the partons are observable in the sense of parton-hadron
duality or deep-inelastic scattering; equivalently, low-energy coloured
vacuum fluctuations cannot propagate into asymptotic states.

Previously, I studied
the induced quantum decoherence and entropy production
in a non-relativistic single-particle model resembling an
electron coupled to the quantized electromagnetic field,
however, with a deliberately enhanced oscillator spectral density
in the infrared. The Feynman-Vernon influence
functional technique for quantum Brownian motion
provided the remarkable result that
in the {\it short-time strong-coupling limit} the model parton
behaves like a {\it classical particle} \cite{I}:
Gaussian parton
wave packets experience {\it friction} and {\it localization}, i.e.
no quantum mechanical spreading, and their coherent {\it superpositions
decohere}. The decoherence process has been shown to
lead to entropy production in this oversimplified parton model.

It seems somewhat more realistic to consider two coupled scalar fields
representing partons and their non-perturbative environment,
respectively. In the functional
Schr\"odinger picture employing Dirac's time-dependent variational
principle, i.e. a non-perturbative method, I derived
a Cornwall-Jackiw-Tomboulis (CJT) type effective action and the
equations of motion for renormalizable interactions \cite{I}. Thus,
analysis of the entropy puzzle in strong interactions
leads to study an observable field
(open subsystem) interacting with a dynamically hidden one
(unobservable environment), i.e. {\it quantum field Brownian motion}.

Summarizing, my point of view is that
partons feel an unobservable
(gluonic) environment, which manifests its strong non-perturbative
interactions
on a short time scale ($\ll 1\;$fm/{\it c}) through
decoherence of suitable partonic pointer states \footnote{In
general, these are not single-particle states but rather
coherent (Gaussian) wave functionals, as constructed in
the second of Refs. \cite{I}.},
their quasi-classical behaviour, and entropy production.
If confirmed in QCD, this will have important consequences
for parton-model applications to complex hadronic
or nuclear reactions. The emergence of
{\it structure functions} from initial-state wave functions
can and will be further studied in this approach.

Instead of representing the formalism and more technical results from
Refs. \cite{I}, I want to demonstrate here in simple quantum
mechanical examples the basic Why and How of the solution to the
entropy puzzle.

Consider a system that can be described in terms of two normalized
discrete basisstates, $|1\rangle$ and $|2\rangle$. Forming a {\it
pure state}, $|\psi\rangle\equiv a_1|1\rangle +a_2|2\rangle$, by a
coherent superposition with amplitudes $a_1\equiv p^{1/2}$ and
$a_2\equiv (1-p)^{1/2}$, the corresponding density matrix,
$\hat{\rho}\equiv |\psi\rangle\langle\psi |$, is
\beq
\rho _{ij}\; =\;
\left ( \begin{array}{cc}
a_1^{\;2} & a_1a_2    \\
a_1a_2    & a_2^{\;2} \\
\end{array} \right )
\;\;\;\longrightarrow\;\;\;
\rho _{ij}^{\;D}\; =\;
\left ( \begin{array}{cc}
1 & 0    \\
0 & 0    \\
\end{array} \right )
\;\;, \eeq{1}
where $\hat{\rho}^D$ is obtained by diagonalization. Note the
off-diagonal {\it interference terms} in $\rho _{ij}$.
Furthermore, observe that $\hat{\rho}^D$ has only one non-vanishing
eigenvalue. Introducing the {\bf von Neumann} or {\bf statistical
entropy},
\beq
S[\hat{\rho}]\;\equiv\; -\;\mbox{Tr}\;\hat{\rho}\;\ln\;\hat{\rho}
\;\;, \eeq{2}
we find $S[\hat{\rho}]=S[\hat{\rho}^D]=-[1\ln 1+0\ln 0]=0$, i.e.
{\it no entropy in a pure state}. Secondly, forming a {\it mixed
state} (ensemble) such that the system is in state $|1\rangle$ with
probability $p$ and in state $|2\rangle$ with probability $1-p$, the
density matrix becomes $\hat{\rho}'=|1\rangle p\langle 1|+|2\rangle
(1-p)\langle 2|$, i.e. a decoherent superposition.
Hence, we obtain
\beq
\rho '_{ij}\; =\;
\left ( \begin{array}{cc}
p & 0   \\
0 & 1-p \\
\end{array} \right )
\;\;. \eeq{3}
The density matrix (\ref{3}) shows {\it no interference terms} and is
diagonal per se.\footnote{Note that $\mbox{Tr}\;\hat{\rho}=
\mbox{Tr}\;\hat{\rho}'=1$: the system is in {\it some} state with
total probability 1.} Then, $S[\hat{\rho}']\equiv
S(p)=-[p\ln p+(1-p)\ln (1-p)]\neq 0$, generally. In fact, $0\leq
S(p)\leq S(1/2)=\ln 2$. Total ignorance about the state of the system
($p=1/2$) corresponds to $\ln 2$ units of entropy in a two-state
system, i.e. 1 {\it bit} of information is lost compared to certainty
about its state ($p=0,1$).

One concludes that {\it entropy production can only occur if
the interference terms of the density matrix representing a more or
less pure state of the observed system decay dynamically}.\footnote{The
argument does not depend on particular physical characteristics of the
system; it holds for the two-state system as well as for
an interacting quantum field.} In a {\it closed system} evolving
unitarily in time, however, there is no way to transform, for
example, $\hat{\rho}$ into $\hat{\rho}'$, see Eqs. (\ref{1}), (\ref{3}).
Only the interaction of the system with an {\it environment}
\cite{I,Zu0,GH},
can have such an effect.
The presence (and integrating out) of
the environment degrees of freedom essentially changes the dynamics of
the observed system. This can lead to the decay of the interference
terms in its density matrix, i.e. {\it environment-induced quantum
decoherence}, which is necessary to increase its impurity and, thus,
to produce entropy.

Next, consider a non-relativistic particle moving in a one-dimensional
double-well potential
presenting the observable subsystem, which is
coupled translationally invariant to
a single environment oscillator. The classical action is
\beq
{\cal S}\; =\;
\int dt\;\left\{\half M\dot{x}^2+\half m\dot{y}^2-\half m\omega ^2
(y-x)^2+\half M\Omega ^2x^2-\frac{1}{4!}\lambda M\Omega ^4x^4\right\}
\;\;. \eeq{4}
For simplicity, let $M=m=\Omega$. Then, properly rescaling
by $\Omega$, one obtains
\beq
{\cal S}\; =\;
\int dt\;\left\{\half \dot{x}^2+\half \dot{y}^2-\half\omega ^2
(y-x)^2+\half x^2-\frac{1}{4!}\lambda x^4\right\}
\;\;, \eeq{5}
in terms of dimensionless quantities
and two coupling constants, $\omega ^2,\lambda\geq 0$. For $\omega =0$
the minima of the doublewell lie at $x_\pm =\pm(\lambda /3!)^{-1/2}$,
at a depth of
$-3/2\lambda$ (the local maximum is zero at
$x=0$). Presently, I want to study the case that the excitation energy of
the x-particle (X) is smaller than the level spacing
$\omega$ of the environment oscillator (Y), which is assumed to be
in its ground state. Starting with a
given initial state of X, I will calculate the time evolution of the
corresponding density matrix $\hat{\rho}_X$ under the influence of
the {\it vacuum fluctuations} of Y. Excited states
of Y contribute only {\it virtually} here;
they cannot become real due to energy conservation.

For illustration,
I choose the metastable initial state of X, when classically
the particle ``rests on top of the hill'' ($x=0$). Quantum mechanically
this can be represented by a minimum uncertainty Gaussian wave packet,
\beq
\psi (x,t=0)\; =\;\pi ^{-1/4}w_0^{-1/2}\mbox{e}^{-\half x^2/w_0^{\;2}}
\;\;, \eeq{6}
with $w_0\ll \lambda ^{-1/2}$.
Also, assume $\omega$ to be sufficiently larger than $3/2\lambda$.

First of all, let the system evolve {\bf classically}. Nothing will
happen. However, any infinitesimal perturbation of the
fine-tuned initial conditions causes X to move ``down the hill'', left
or right (L or R), dragging Y along. There is local chaos in the sense
of extreme sensitivity to the initial conditions at $x,y\approx 0$;
arbitrarily small
uncertainties in the initial conditions lead to a {\it loss of
predictability}. For an {\it ensemble of initial conditions}
X switches with probabilities
$p_L(t')$ and $p_R(t')=1-p_L(t')$ between L and R, respectively,
if at least
one trajectory passes $x=0$ in a certain interval
$[t'-\epsilon ,t'+\epsilon ]$. This corresponds to a loss of
information about the actual binary decision ``either L or R'' and
an entropy $S_X=-\sum _{i=L,R}p_i\ln p_i$.\footnote{If the ensemble
of initial conditions is constrained to preserve the reflection
symmetry of the action, Eq. (\ref{5}),
then $p_L(t')=p_R(t')=\half$ and $S_X(t')=\ln 2$ stay constant.}
Note that $S_X$ or $p_{L,R}$
are strongly conditional (``fine-grained'') quantities.
In distinction, the usual classical
entropy is calculated after ``coarse graining'',
i.e. by constructing a local probability density $f(t)$
in the phase space of X related to
the ensemble average over initial conditions,
$S_{c.g.}(t)\equiv\int dxdp\; f(t)\ln f(t)+C$. A chaotic loss of
predictability from strongly diverging trajectories in phase space
causes $S_{c.g.}$ to increase: as time passes, more and more
cells of the coarse graining contribute $-$ effectively, the
phase space volume occupied by the ensemble grows (without violating
Liouville's theorem).

In conclusion, {\it in a classical system, be it chaotic or not (with
or without coarse graining), entropy
can only be produced IF there is a physically relevant ensemble
of initial conditions}. Thus, one cannot explain altogether classically
entropy production or thermalization in a single high-multiplicity
event in strong interactions.

Secondly, let the system evolve {\bf quantum mechanically}. To begin
with, let there be {\it no} coupling to the environment ($\omega =0$).
Even with the fine-tuned initial condition, Eq. (\ref{6}),
the amplitude $\psi$
to find X at a particular space-time point begins to flow ``down the
hill'' symmetrically (L and R) due to the quantum spreading of the
wave packet. For a free particle $w_0\;\rightarrow\; w(t)=(w_0^2+
w_0^{-2}t^2)^{1/2}$ (for $M=1$); here one
expects an accelerated spreading ``downhill'', cf. Eq. (\ref{12}) below.
 The related probability density $|\psi |^2$ also evolves and
stays symmetric; generally, it {\it cannot}
be simulated by the classical evolution starting with an ensemble
of initial conditions due to the absence of quantum interference between
classical trajectories. In any case,
the system remains in a {\it pure quantum
state}. The density matrix is $\hat{\rho}_X(t)=|\psi (t)\rangle\langle
\psi (t)|$, $|\psi (t)\rangle =\exp [-i\hat{\mbox{H}}_0t]|\psi (0)
\rangle$, where $\hat{\mbox{H}}_0$ is the Hamiltonian of X from
Eq. (\ref{5}) with $\omega =0$. Therefore, $S[\hat{\rho}_X(t)]=
-1\ln 1=0$, cf. Eq. (\ref{2}). Quantum mechanically one knows
everything there is to know about a {\it closed system} (X),
given any
pure initial state and its Hamiltonian,
which consistently yields $S[\hat{\rho}_X]=0$.
Even with an ensemble of initial states, i.e.
an impure density matrix $\hat{\rho}_X(0)$, there is {\it no
entropy production}, since $S[\hat{\rho}_X(t)]=S[
\exp (-i\hat{\mbox{H}}_0t)\hat{\rho}_X(0)
\exp (+i\hat{\mbox{H}}_0t)]=S[\hat{\rho}_X(0)]$ stays
constant.\footnote{The
unitary (time evolution) transformation does not change the
eigenvalues of $\hat{\rho}_X$. Thus, the statistical entropy, Eq.
(\ref{2}), cannot possibly show a sign of classical chaos in a closed
system.}

The situation changes completely if the subsystem (X)
evolves quantum mechanically coupled to the {\it vacuum fluctuations and
virtual excitations of the environment} (Y). With the above
assumptions the initial density matrix of the total system is:
\beq
\hat{\rho}(t=0)\;\equiv\;\hat{\rho}_X(0)\otimes\hat{\rho}_Y(0)
\;\;, \eeq{7}
with matrix elements $\rho _X(x,x';0)=\pi ^{-1/2}w_0^{-1}\exp [-
\half (x^2+x'^2)/w_0^{\;2}]$ and
$\rho _Y(y,y';0)=$ $(\omega /\pi )^{1/2}\exp [-
\half\omega (y^2+y'^2)]$. The time evolution of the density matrix
of the observable subsystem, $\hat{\rho}_X(t)=\mbox{Tr}_Y
\hat{\rho}(t)$, can be calculated with the Feynman-Vernon influence
functional technique; I will make use of general results obtained in
the first of Refs. \cite{I}.
The idea is to derive a propagator for $\hat{\rho}_X$, which
incorporates the influence of the environment degrees of freedom (Y)
exactly. This can be achieved, since Y and its coupling to X are
at most quadratic in coordinates and momenta, see Eq. (\ref{5}).

It should be remarked that the {\it final state} of the environment
is not specified; presently, it may contain virtual excitations of
Y.\footnote{As a corollary to the {\it Schmidt decomposition} \cite{I}
it is easy to prove that starting with an overall pure state of the
complex system, cf. Eq. (\ref{7}), IF the final state of the
environment is a pure state, THEN the observable subsystem ends up
in a pure state too (without entropy
production).} The relevance of this for a
high-multiplicity hadronic (or nuclear) reaction is the
following: Even
though the QCD vacuum ``far away'' conforms to the usual one
before and after, the additionally produced secondary hadrons
all require a dressing of their valence quarks by localized virtual
excitations of the vacuum or environment, which obviously makes an
essential difference as compared to the initial state.

Presently, the resulting density matrix $\hat{\rho}_X(t)$ is
(cf. also the first of Refs. \cite{I}):
\beqar
\rho _X(z_-,z_+,t)&=&\pi ^{-1/2}w^{-1}(t)\;
\mbox{e} ^{\textstyle{-[z_+-v(t)t]^2/w^2(t)}}
\nonumber \\ [1ex]
&\;&\times\;
\mbox{e} ^{\textstyle{-z_-^{\;2}\{ C+\quarter w_0^{\;2}c^2-d^{-2}
[B+\half w_0^{\;2}bc]^2/w^2(t)\} }} \nonumber \\ [1ex]
&\;&\times\;\mbox{e} ^{\textstyle{iz_-\{ (az_+-2d^{-1}
[B+\half w_0^{\;2}bc][z_+-v(t)t]/w^2(t)\} }}
\;\;, \eeqar{8}
with the effective velocity
$v(t)=0$ for the zero-momentum initial wave packet, the effective width
$w(t)\equiv 2\xi |d|^{-1}$, $\xi\equiv (A+\quarter w_0^{-2}+\quarter
w_0^{\;2}b^2)^{1/2}$,
and with rather complicated time-dependent coefficients
$A,\; B,\; C,\; a,\; b,\; c,\; d$, to be discussed elsewhere;
the coordinates in Eq. (\ref{8})
are $z_-\equiv x-x'$ and $z_+\equiv\half (x+x')$ in terms of ordinary
one-dimensional ones. Since we are particularly interested
in the decoherence process and entropy production, we
consider only the simplest {\it off-diagonal density matrix elements}
here,
$\rho _X(x,x'=-x,t)=\rho _X(z_-=2x,z_+=0,t)$.
They can be directly related to the
{\bf linear entropy} produced in the observable subsystem (X):
\beqar
S^{lin}&\equiv&\mbox{Tr}\; [\hat{\rho}_X-\hat{\rho}_X^{\;2}]\; =\;
1-\int _{-\infty}^{\infty}dz_-\int _{-\infty}^{\infty}dz_+\;
\rho _X(z_-,z_+,t)\;\rho _X(-z_-,z_+,t) \nonumber \\
&=&1-\half c_1^{-1/2}w^{-1}
\;\;, \eeqar{9}
with
$c_1\;\equiv\;C+\quarter w_0^{\;2}c^2-(B+\half w_0^{\;2}bc)^2/(dw)^2$ and
 independently of the initial wave packet momentum ($p=0$ at present).
Thus, inserting (\ref{9}) into (\ref{8}), one obtains:
\beqar
&\;&\rho _X(x,-x,t)\; =\;\pi ^{-1/2}w^{-1}(t)\;\exp\left\{ -x^2w^{-2}(t)
[1-S^{lin}(t)]^{-2}\right\} \;\;, \label{10} \\[2ex]
&\;&\int _{-\infty}^{\infty}dx\;\rho _X(x,-x,t)\; =\; 1-S^{lin}(t)\;
\geq\;\mbox{e}^{-S(t)}
\;\;. \eeqar{11}
The inequality results from the fact that the linear
entropy
provides a {\it lower bound} for the relevant statistical entropy,
cf. Eq. (\ref{2}), as shown in \cite{I}.
Note that Eqs. (\ref{9})$-$(\ref{11}) are completely independent of
the time-dependent functions entering there, which are specific for a
particular dynamical system. They are based, however, on the
Gaussian structure of the subsystem density matrix, Eq. (\ref{8}).

At this point the attentive reader might wonder what happened to the
non-linear interaction $\propto\lambda x^4$ of the double-well potential,
 see Eq. (\ref{5}). Of course, it cannot be treated exactly. I employed
a mean-field-type approximation, replacing $\frac{1}{4!}\lambda x^4$ by
$\half\lambda\langle x^2\rangle x^2\equiv\half\Lambda ^2(t)x^2
$. As long as one studies only the initial time-evolution over short
periods, as compared to the
time a classical particle would need to ``roll down the hill'',
one may even set $\Lambda\approx 0$. For the following qualitative
considerations, $\Lambda$ plays the role of an adiabatically changing
parameter.
However, a more accurate approximation is necessary (and feasible) to
follow the truly long-time quasi-periodic motions of the system.
Then, one expects periods of increasing decoherence and entropy
production, cf. below, followed by periods of
{\it quantum revival} in the observed subsystem.
The more complex the environment becomes,
the more unlikely quantum revivals will be, since the total system
including the environment finds more and more ways to evolve before
a reconstruction of the subsystem initial-state wave
function.\footnote{Such effects have been experimentally
observed in even simpler systems involving a two-state subsystem of
one Rydberg atom coupled to a single mode of the electromagnetic
field \cite{Walther}.}

To begin with, it can be checked
explicitly that there is no entropy
production for a vanishing coupling to the environment, $S_{\omega =0}
^{lin}(t)=0$, cf. Eq. (\ref{9}). Next, calculating the effective
width in the {\it long-time limit} \footnote{Here, hyperbolic functions
dominate over trigonometric ones in the time-dependent
coefficients in Eq. (\ref{8}) for times such that a classical
particle would still be ``rolling down the hill'' of the potential;
this restriction is presently assumed for simplicity.}, one finds:
\beq
w(t)\; =\; \left (w_0^{\;2}+w_0^{-2}f_-^{-2}+\omega ^3f_-^{-2}
[f_-^{\;2}+\omega ^2]^{-1}
\right )^{1/2}\frac{f_-^{\;2}+\omega ^2}{f_-^{\;2}+f_+^{\;2}}\;
\exp\; t_-
\;\;\;, \eeq{12}
with $t_-\equiv f_-t$, $f_\pm\equiv [\pm\half\omega _+^{\;2}+
(\quarter\omega _+^{\;4}-\omega ^2\omega _-^{\;2})^{1/2}]^{1/2}$,
and $\omega _\pm^{\;2}\equiv\pm\omega ^2+\Lambda ^2-1$. Assuming a
sufficiently small coupling, $\omega ^2<1$, note that $\omega _\pm
^{\;2}$ is {\it negative} as long as $\Lambda ^2(t)<1-\omega ^2$.
Thus, the {\it width grows
exponentially} with an {\it effective Lyapunov exponent} $f_-$.
For vanishing coupling to the environment,
it reduces to the classical Lyapunov exponent of the inverted
oscillator, $f_-^{\omega =0}
=(1-\Lambda ^2)^{1/2}$, while the width becomes
$w_{\omega =0}(t)=(w_0^{\;2}+
[w_0f_-^{\omega =0}]^{-2})^{1/2}\exp (f_-^{\omega =0}t)$. This
suggests quite generally that the time-dependent widths of suitable
(Gaussian) wave packets may serve as ``quantum indicators'' of chaotic
behaviour in the corresponding classical system.

It is remarkable how the Lyapunov exponent reflects the
dynamics: as the wave packet spreads ``downhill'', $\Lambda ^2(t)\propto
\langle x^2\rangle$ increases until $f_-$ reaches zero (becoming
purely imaginary afterwards), when $\omega _-=0$. At this point the
behaviour becomes regular in the sense of being governed by harmonic
motions close to the minima of the double-well potential with a
correspondingly milder time-dependence of the width (cf. the
model studied in the first of Refs. \cite{I}).

The second dynamical time scale $f_+^{-1}$ always stays real.
It is relevant for certain non-Markovian
effects generated by the interaction with the environment ($f_+
^{\omega =0}=0$). These become clearly
visible in the entropy evaluated in the same limit as Eq. (\ref{12}).
Using $c_1=C+\mbox{O}(w_0^{\;2})$, one obtains in leading order:
\beqar
&\;&S^{lin}(t)\; =\; 1-w^{-1}(t)\;\frac{f_-^{\;2}+\omega ^2}
{\omega ^{3/2}f_-} \;\times \label{13} \\
&\;&
\left [[(\frac{f_-[f_+^{\;2}-\omega ^2]}{f_+[f_-^{\;2}+f_+^{\;2}}
           +\frac{f_+[f_-^{\;2}+\omega ^2]}{f_-[f_-^{\;2}+f_+^{\;2}})
\sin t_+ +\frac{2\omega ^2\cos t_+}{f_-^{\;2}+f_+^{\;2}}]^2
+\omega ^2[\frac{\sin t_+}{f_+} -\frac{\cos t_+}{f_-}]^2\right ]^{-1/2}
, \nonumber \eeqar{13.1}
with $t_+\equiv f_+t$. Thus, the linear entropy approaches
exponentially its saturation value 1 on the time scale set by the
Lyapunov exponent, see Eq. (\ref{12}),
and the von Neumann entropy grows exponentially according to
eq. (\ref{11}), at least as fast. Note that the periodic function
multiplying $w^{-1}(t)$ in Eq. (\ref{13}) is approximately
$\propto\omega ^{-1/2}[1+2^{1/2}+\sin ^2(2^{1/4}\omega t)]^{-1/2}$ for
$f_-\approx f_+$ and $\propto\omega ^{-1/2}[1+\cos ^2(2^{1/2}\omega t)]
^{-1/2}$ for $f_-\approx 0$; i.e. it persists qualitatively
even until the effective Lyapunov exponent becomes imaginary, when
the stabilizing effect of the $x^4$-term in the potential is felt.

To summarize, in the above specified long-time limit and for the
chosen initial conditions, Eq. (\ref{7}), one obtains the observable
subsystem (X) density matrix,
\beq
\rho _X(z_-,z_+,t)\; =\;\pi ^{-1/2}w^{-1}(t)\;
\mbox{e} ^{\textstyle{-\{ z_+^{\;2}
                      +\quarter z_-^{\;2}[1-S^{lin}(t)]^{-2} \}
w^{-2}(t)+iz_-z_+f_-}}
\;\;. \eeq{14}
Even though this density matrix describes the exponential entropy
production and its diagonal matrix elements with $z_-=0$
grow rapidly, apart from the overall normalization factor, the
(simplest) off-diagonal matrix elements ($z_+=0$) do {\it not} really
decay here as in usual models of quantum decoherence \cite{I,Zu0}.
This is no surprise in view of the ``poor environment'' considered at
present, which has only one degree of freedom frozen in its ground
state (modulo virtual excitations). Loosely speaking, it is unable
to accommodate all the phase information contained in the off-diagonal
density matrix elements of the subsystem.

{\it In conclusion, a strong observable entropy production in a quantum
system, which shows a chaotic behaviour in the classical limit with
exponentially growing modes, requires only a minimal decohering effect
due to an environment of vacuum fluctuations
coupled to it from a higher energy scale.}

In particular, complementary to previous studies \cite{Zu0,Zu1}, one
observes here that the environment does not necessarily have to be at
any finite temperature for this effect of {\it partial decoherence}
to work. Furthermore, it should be realized that
the Schmidt decomposition reveals the remarkable fact that
the density submatrices of ``subsystem'' and ``environment'' always
have identical non-zero eigenvalues \cite{I}. Thus, from the point of
view of calculating the entropy, Eq. (\ref{2}), their roles can be
interchanged and it is a matter of practicability to decide which part
of the total Hilbert space is integrated out to find the entropy of the
{\it physically observed subsystem}.
The environment-induced quantum decoherence and its
relevance for entropy production
have presenly been illustrated by an elementary example, which, however,
points out to interesting consequences for the quantum evolution of
classically chaotic non-linear field theories.

\vskip .15cm
I thank P. Carruthers, M. Danos, N. E. Mavromatos, B. M\"uller, O.
Bertolami, and J. Rafelski
for stimulating criticism and helpful discussions.

\end{document}